\documentclass[runningheads]{llncs}

\usepackage{graphicx}
\usepackage{xcolor} 
\usepackage{paralist}
\usepackage{listings}
\usepackage{float}
\usepackage[colorlinks=true,hidelinks,breaklinks=true]{hyperref}
\usepackage{epsfig}
\usepackage{caption}
\usepackage{subcaption}
\usepackage{todonotes} 

\newcommand{\isc}{ISC}

\newcommand{\sql}{SQL}
\newcommand{\xes}{XES}
\newcommand{\csv}{CSV}
\newcommand{\jar}{JAR}

\newcommand{\constraint}[1]{\emph{``#1''}} 

\newcommand{\qall}{Q_{\mathrm{case}}}
\newcommand{\qv}{Q_{\mathrm{viol}}}
\newcommand{\qs}{Q_{\mathrm{sat}}}
\newcommand{\qpendv}{Q_{\mathrm{viol-pending}}}
\newcommand{\qpends}{Q_{\mathrm{sat-pending}}}
\newcommand{\qpermv}{Q_{\mathrm{viol-perm}}}
\newcommand{\qperms}{Q_{\mathrm{sat-perm}}}

\definecolor{codegreen}{rgb}{0,0.6,0}
\definecolor{codegray}{rgb}{0.5,0.5,0.5}
\definecolor{codepurple}{rgb}{0.58,0,0.82}
\definecolor{backcolour}{rgb}{0.95,0.95,0.92}

\lstdefinestyle{mystyle}{
    commentstyle=\it\color{codegreen},
    keywordstyle=\color{magenta},
    numberstyle=\tiny\color{codegray},
    stringstyle=\color{codegray},
    basicstyle=\ttfamily\footnotesize,
    breakatwhitespace=false,
    breaklines=true,
    captionpos=b,
    keepspaces=true,
    numbersep=5pt,
    showspaces=false,
    showstringspaces=false,
    showtabs=false,
    tabsize=2,
    deletekeywords={Timestamp}
}

\begin{document}

\title{What Can Database Query Processing Do for Instance-Spanning Constraints?}
\titlerunning{Query Processing for \isc}

\author{Heba Aamer \inst{1}\orcidID{0000-0003-0460-8534} \and
Marco Montali \inst{2}\orcidID{0000-0002-8021-3430} \and
Jan Van den Bussche \inst{1}\orcidID{0000-0003-0072-3252}}

\authorrunning{Aamer et al.}

\institute{Hasselt University, Belgium \\
\email{\{heba.mohamed,jan.vandenbussche\}@uhasselt.be}\\ \and
Free University of Bozen-Bolzano, Italy \\
\email{montali@inf.unibz.it}}

\maketitle

\begin{abstract}
In the last decade, the term \emph{instance-spanning constraint} has been introduced
in the process mining field to refer to constraints that span multiple process
instances of one or several processes. Of particular relevance, in this setting, is checking whether process executions comply with constraints of interest, which at runtime calls for suitable monitoring techniques.  Even though event data are often stored in some sort of database,
there is a lack of database-oriented approaches to tackle compliance checking and monitoring of (instance-spanning) constraints.  In this paper, we fill this gap by showing how well-established technology from database query processing can be effectively used for this purpose.  We
propose to define an instance-spanning constraint
through an ensemble of \emph{four database queries} that retrieve the satisfying,
violating, pending-satisfying, and pending-violating cases of the constraint.
In this context, the problem of compliance monitoring then becomes an
application of techniques for incremental view maintenance, which is
well-developed in database query processing.  In this paper, we argue for
our approach in detail, and, as a proof of concept, present an experimental
validation using the DBToaster incremental database query engine.

\end{abstract}

\begin{minipage}{0.5\textwidth}
  \hfill
\end{minipage}
\begin{minipage}{0.45\textwidth}
  \begin{description}
    \item [Q:] What's in a constraint?
    \item [A:] Two (or four) database queries!
  \end{description}
\end{minipage}

\section{Introduction}
\label{sec:intro}

\constraint{Paying for something purchased online cannot happen after receiving
it}, \constraint{The average time for a package to be delivered after purchase
is between two and five days}, and \constraint{The same shipping car can be
used for delivering packages at most seven times per day} are various examples
of constraints that are posed over business processes.
These constraints can be very general and can refer to a variety of requirements~\cite{LyMMRA15}.  Non-compliance of certain constraints can be very costly
and risky, so compliance checking\footnote{One should differentiate between the
problems of \emph{verification} and \emph{compliance checking}.  Our focus is
on compliance checking: checking properties of execution logs.
On the other hand, in verification, one seeks to determine whether all possible
executions of some given process model satisfy some property.  The kind of
constraints we are dealing with in this paper are typically quite expressive,
so that verification would be undecidable and one needs to resort to compliance
checking.
There is also a third problem, \emph{conformance checking}~\cite{Aalst12},
where we check that a given execution follows a given process model.  This
problem is outside the scope of this paper, although, formally speaking,
conformance checking could be viewed as a kind of compliance checking.}
and monitoring are of utmost importance to the enterprise~\cite{WinterSR20}.

Constraints can be very simple in terms of their scope, i.e., the process
instances they involve, and the conditions they impose such as
\constraint{Conducting a patient's surgery must be preceded by examining the
patient} or \constraint{Paying for something purchased online cannot happen
after receiving it}.  Those are examples of constraints to be enforced on activity instances belonging to
the same process instance.  This type of constraint is often referred to as
\emph{intra-instance}~\cite{WarnerA06,WinterSR20}.  On the other
hand, there are constraints that can be much more complex, both in their scope
and in the conditions they impose.
Specifically, constraints where the scope spans multiple process
instances, or combinations of entities involved in multiple process instance,
have been referred to as \emph{inter-instance}~\cite{WarnerA06,MontaliMCMA13}, or, more
recently, \emph{instance-spanning} constraint
(\isc)~\cite{FdhilaGRMI16,Rinderle-MaGFMI16}.  \constraint{The same shipping
car can be used for delivering packages at most seven times per day} and
\constraint{Packages that are delivered to the same neighbourhood on the same
day must be delivered by the same shipping car} are examples of \isc.

\begin{figure}
     \centering
     \begin{subfigure}[t]{0.34\textwidth}
         \centering
         \includegraphics[width=\textwidth]{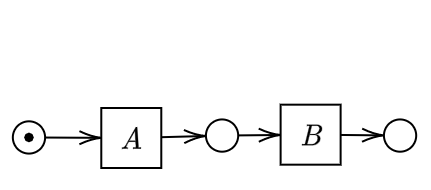}
         \caption{}
         \label{fig:p1model1}
     \end{subfigure}
     \hfill
     \begin{subfigure}[t]{0.64\textwidth}
         \centering
         \includegraphics[width=\textwidth]{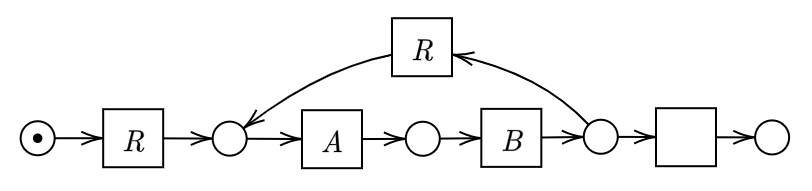}
         \caption{}
         \label{fig:p1model2}
     \end{subfigure}
\caption{Consider the process $P_{AB}$ in Figure~\ref{fig:p1model1}.
\constraint{There can be at most three orders per customer} is an example of
an \isc~when posed against multiple instances of $P_{AB}$.  On the other hand,
when the same constraint is posed against the iterative model in
Figure~\ref{fig:p1model2}, then it would be an intra-instance constraint.
Note that $R$ added in Figure~\ref{fig:p1model2} would resemble a
receive order activity.}
\label{fig:p1-models}
\end{figure}

\begin{figure}
     \centering
     \begin{subfigure}[t]{0.35\textwidth}
         \centering
         \includegraphics[width=\textwidth]{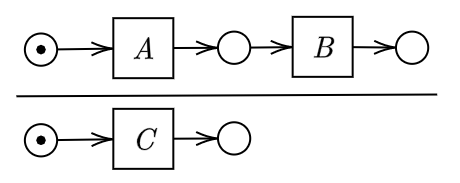}
         \caption{}
         \label{fig:p2model1}
     \end{subfigure}
     \hfill
     \begin{subfigure}[t]{0.6\textwidth}
         \centering
         \includegraphics[width=\textwidth]{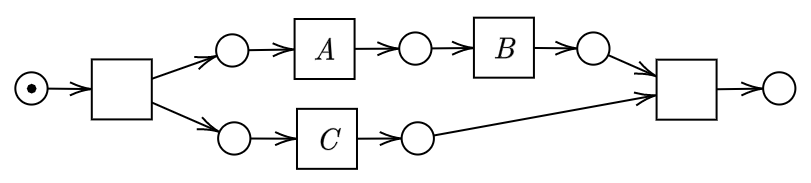}
         \caption{}
         \label{fig:p2model2}
     \end{subfigure}
\caption{Consider the two separate processes $P_{AB}$ and $P_{C}$ in
Figure~\ref{fig:p2model1}.  \constraint{For every instance of $P_{AB}$, an
instance of $P_{C}$ must be instantiated for the same customer} is an example
of \isc~that relates instances of the two processes based on a
common attribute.  On the other hand, when the two processes are subprocesses
of a single process as in Figure~\ref{fig:p2model2}, the constraint would be
an intra-instance constraint.}
\label{fig:p2-models}
\end{figure}

It should be noted, however, that whether a constraint is intra-instance or
instance-spanning is a relative matter; it depends on the design of the process
model.  Indeed, in general, a single process may require sophisticated control-flow structures involving iterations and  multi-instance activities.  Figures~\ref{fig:p1-models} and~\ref{fig:p2-models} give
simple illustrations of the relative nature of ``intra'' versus ``inter''
instance.  Thus, while our focus is on studying \isc, similar features would be required when checking intra-instance constraints on such complex processes.  In what
follows, we will hence just talk about (process) \emph{constraints} in general.

Constraints must be checked against execution logs, which are files or
databases holding data about past and current executions of all process
instances in the enterprise.  Two types of compliance checking are commonly
distinguished:
\begin{description}
  \item[Post-mortem checking] targets only full (completed) executions on a
  historical log.
  \item[Compliance monitoring] checks the execution of the currently running
  process instances, for a live log.\footnote{Of course, in principle, post
  mortem checking can also be performed within a live log.}
\end{description}

There is a striking similarity between the problem of compliance monitoring
and the problem of \emph{incremental view maintenance}, a well-researched
problem in databases~\cite{GuptaMS93,GuptaM95,GuptaM99Book,ChirkovaY12Book,KennedyAK11,KochAKNNLS14}.
There, a \emph{view} is the materialized result of a (possibly complex) query
posed against a database.  The problem of view maintenance is then to keep the
view consistent with its  definition under changes to the database.  In
general, these changes may be CRUD operations such as in particular insertions, deletions, or updates. This is perfectly in line with the execution of a process, where events witness the execution of tasks that, in turn, are typically associated to CRUD operations used to persist relevant event data in an underlying storage.

\emph{In this paper, we put forward the idea that incremental view maintenance is
applicable to do compliance monitoring.}  To do so, we need to answer three
questions:
\begin{inparaenum}[(1)]
\item what is the database?
\item What are the updates?
\item What is the query?
\end{inparaenum}

The first two questions are easily answered: the log is the database, and events trigger insertions to the log to leave a trace about their occurrence. In this context, only insertion operations are thus used, to append the occurrence to an event to those occurred before. Every insertion, triggered by the execution of some activity instance, stores the corresponding event data in the database, including the timestamp of the event and which data payload it carries.

What is then the query? To answer this question, we first need to indicate which dimensions we want to tackle when expressing constraints. Given the nature of \isc, we want to comprehensively tackle multi-perspective constraints dealing with several cases and their control-flow, time, and data dimensions. Instead of defining a specific constraint language that can accommodate such different perspectives, we directly employ full-fledged SQL for the purpose. Hence, a constraint is expressed as a query or, more precisely, an ensemble of queries, the number of which depends on whether compliance has to be assessed post-mortem or at runtime. In post-mortem checking, a constraint is expressed as a pair $(\qall, \qv)$ of two queries:
\begin{compactitem}
      \item $\qall$ defines the ``scope" of the constraint -- it returns the
      set of cases to which the constraint applies;
      \item $\qv$ returns the subset of cases that violate the constraint.
\end{compactitem}
At runtime, we take inspiration from previous works in monitoring processes and temporal logic specifications \cite{BaLS11,MMWA11,DDGM14}, and consider that each constraint may be, in principle, in one of four possible states: currently satisfied (resp., currently violated), that is, satisfied (resp., violated) by the current event data, but with a possible evolution of the system that will lead to violation (resp., satisfaction); permanently satisfied (resp., permanently violated), that is, satisfied (resp., violated) by the current event data, and staying in that state no matter which further events will occur in the future. For well-studied languages only tackling the control-flow dimension, such as variants of linear temporal logics over finite traces, such states can all be automatically characterized starting from a single formula formalizing the constraint of interest \cite{DDMM22}. This is not the case for richer languages tackling also the data dimension, as in this setting reasoning on future continuations is in general undecidable \cite{Del09,CDMP22}. We therefore opt for a pragmatic approach where constraint states are manually identified by the user through dedicated queries, as in  \cite{MontaliMCMA13,CaMC19}. In particular, a monitored constraint comes with an ensemble of four queries:
$(\qall, \qpermv, \qpendv, \qpends)$
, where:
   \begin{compactitem}
     \item $\qall$ is as before;
     \item $\qpendv$ and $\qpends$ return the ``pending" cases that, respectively, violate and satisfy the constraint at present, but for which upon acquisition of new events, their status may change.
     \item $\qpermv$ returns permanent violations, i.e., those cases that irrevocably violate the constraint, that is, for which the constraint is currently violated and will stay so no matter which further events are collected.
   \end{compactitem}

To monitor constraints, we have used the system DBToaster for incremental query
processing~\cite{KennedyAK11,KochAKNNLS14,DBToasterWeb} in a
proof-of-concept experiment.  We monitor a number of realistic constraints on
experimental data taken from the work by Winter et al.\ \cite{WinterSR20}.
We will present multiple examples demonstrating our approach in
Sections~\ref{sec:examples} and~\ref{sec:monitor} of the paper.

Importantly, while we employ here the de-facto standard query language in databases,
\sql, any other general data model (capable of suitably representing
execution logs) with a sufficiently expressive declarative query language would
do as well.  Examples are the RDF data model with SPARQL, or graph databases
with Cypher.  It should be noted, however, that incremental query processing is
the most advanced for \sql. Indeed, relational database management systems are
still the most mature database technology in development since the 1970s.

The rest of the paper is organized as follows.  In Section~\ref{sec:examples},
we formalize our approach, discuss some examples of constraints and express
them as \sql~queries.  In Section~\ref{sec:monitor}, we elaborate on the
problem of compliance monitoring.  In Section~\ref{sec:exp}, we present the
experimental results.  In Section~\ref{sec:seqdlog}, we discuss query language
extensions for sequences that can be useful for an approach.  We conclude in
Section~\ref{sec:conc}.

\section{Post-mortem Analysis by Queries}
\label{sec:examples}
We capture a constraint as a query that returns the set of cases incurring in a violation.

\begin{definition}[Constraint, Post-mortem Variant]
A \emph{constraint} $C$ is a pair $(\qall, \qv)$ of queries
where $\qall$ is a \emph{scoping query} that returns all the cases subject to the constraint $C$, while
$\qv$ is a \emph{violation detection query} that returns the violating cases such that $\qv$ is always a subset of $\qall$.
\end{definition}

This definition settles our approach for post-mortem checking.  It is simply an
application of query answering, where the queries are asked against a
database instance (representing the execution log) that consists only of
completed process instances.
In that case, when a tuple $t \in \qall \setminus \qv$, then $t$ represents
a case that satisfies the constraint (i.e., $t \in \qs$).

\begin{remark}
Note that an equivalent approach is to represent the constraint as the pair of
queries $(\qall, \qs)$ instead.  The two approaches are interchangeable since
$\qs$ can be defined in \sql~as follows (assuming that both $\qall$ and $\qv$
are materialized):
\begin{lstlisting}[language=SQL,style=mystyle]
( SELECT * FROM Q_case )
EXCEPT
( SELECT * FROM Q_viol ); 
\end{lstlisting}
\end{remark}

\begin{figure}
  \centering
\includegraphics[scale=0.32]{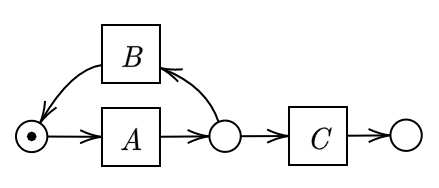}
\caption{\label{fig:pABAC} A process model of an example process.}
\end{figure}

\begin{example}
For an example of a constraint that its $\qs$ query is defined easier than its
$\qv$ query, consider the constraint \constraint{Activity B must be executed
at least once in any process instance.} that is imposed over the process model
given in Figure~\ref{fig:pABAC}.  In this example, defining $\qv$ is more
complicated as it requires negation.  On the other hand, $\qs$ is a simple
existentially quantified statement.
\end{example}

Guaranteeing that, for a constraint $(\qall, \qv)$, query $\qv$ always returns a subset of $\qall$ is under the responsibility of the modeler. One way to ensure this is to write $\qv$ as a query that takes $\qall$ and extends it with a filter to identify violations; however, alternative formulations may be preferred for readability and/or performance needs.

\subsection{Database Schema}

We note that the structure of the database schema representing the data of the
execution log and how to get a database instance with the data are not issues
that we address in this paper.  These problems are orthogonal to what we
discuss in this paper.  In the work by de Murillas et al. \cite{MurillasRA19},
they showed how to \emph{automatically} extract, transform, and load the log's
data from scattered sources into a database instance.  In the same work, they
devised a meta model that structures the database into a specific schema that
is easily queried.

Thus, in our work, we assume that we can have a suitable database schema to
work with.  However, we will not be assuming the schema suggested by de
Murillas et al. as it is very comprehensive, also integrating issues such as
versioning and provenance.  For our purposes of giving illustrating examples,
we will assume the following two relations in our database:
\begin{itemize}
  \item A main $\mathtt{Log}$ relation that has the following schema

  $$\mathtt{(CaseId, EventId, ActivityLabel, Timestamp, Lifecycle)}$$

  The $\mathtt{ActivityLabel}$ and $\mathtt{Timestamp}$ attributes are mandatory
  when working with (instance-spanning) constraints~\cite{WinterSR20}.  The
  $\mathtt{Lifecycle}$ attribute describes the \emph{lifecycle transition} of
  an event.  This is useful when the events can span a time interval which is
  typical in the constraints checking \emph{concurrent} execution of
  activities.  All of those attributes are parts of the XES standard
  extensions~\cite{XESLink}.

  \item An auxiliary $\mathtt{EventData}$ relation that contains the extra
  information of the logged events.  The attributes of this relation are not
  fixed and they (depend on the application) change depending on the data,
  however, the key of this relation is the pair $\mathtt{(EventId, Lifecycle)}$.
\end{itemize}

\begin{remark}
An alternative approach to define the schema of $\mathtt{EventData}$ relation
is by following a semi-structured approach.  In that approach, the schema is
fixed to be $\mathtt{(EventId, Lifecycle, Attribute, Value)}$, where
$\mathtt{Attribute}$ could be the name of the attribute, while
$\mathtt{Value}$ is its value for that event.
\end{remark}

\subsection{Examples}

In the following examples, we assume that the relation $\mathtt{EventData}$
has the following schema $\mathtt{(EventId, Lifecycle, PackageId, CarId)}$.
We also assume that in our processes, we have two activities with the labels
``purchase package'' and ``deliver package''.

\begin{example}[Same Shipping Car Constraint]\label{ex:same_car1}
Consider the constraint \constraint{The same shipping car can be used for
delivering packages at most seven times per day}.  As we have mentioned before,
we have a great flexibility in defining what a violation is (in other words,
what is the scope of the constraint).  One possibility is to define the cases
to be tuples $\mathtt{(CarId, Day)}$.  Following this view, the constraint can
be represented by the following pair of queries:

\begin{lstlisting}[language=SQL,style=mystyle]
-- Q_case

SELECT e.CarId, DATE(l.Timestamp)
FROM Log l, EventData e
WHERE l.EventId=e.EventId AND
      l.ActivityLabel='deliver package' AND
      l.Lifecycle='complete' AND e.Lifecycle='complete'
GROUP BY e.CarId, DATE(l.Timestamp);

-- Q_viol

SELECT e.CarId, DATE(l.Timestamp)
FROM Log l, EventData e
WHERE l.EventId=e.EventId AND
      l.ActivityLabel='deliver package' AND
      l.Lifecycle='complete' AND e.Lifecycle='complete'  
GROUP BY e.CarId, DATE(l.Timestamp)
HAVING COUNT(e.PackageId) > 7;
\end{lstlisting}

A less fine-grained scope: only having $\mathtt{CarId}$.  An even more
fine-grained scope: having tuples of $\mathtt{(CarId, Day, CountOfDeliveries)}$
as our cases.

\begin{lstlisting}[language=SQL,style=mystyle]
-- Q_case

SELECT e.CarId, DATE(l.Timestamp), COUNT(e.PackageId)
FROM Log l, EventData e
WHERE l.EventId=e.EventId AND
      l.ActivityLabel='deliver package' AND
      l.Lifecycle='complete' AND e.Lifecycle='complete'
GROUP BY e.CarId, DATE(l.Timestamp);

-- Q_viol

SELECT e.CarId, DATE(l.Timestamp), COUNT(e.PackageId)
FROM Log l, EventData e
WHERE l.EventId=e.EventId AND
      l.ActivityLabel='deliver package' AND
      l.Lifecycle='complete' AND e.Lifecycle='complete'  
GROUP BY e.CarId, DATE(l.Timestamp)
HAVING COUNT(e.PackageId) > 7;
\end{lstlisting}
\end{example}

Example~\ref{ex:same_car1} demonstrates possible queries that define an
instance-spanning constraint.  To show the uniformity of our approach, the
following is an example of an intra-instance constraint.

\begin{example}[Average Shipping Time Constraint]\label{ex:avg_deliver}
Consider the constraint \constraint{The average time for a package to be
delivered after purchase is between two and five days}.  In what follows,
we consider a case to be a package identifier.

\begin{lstlisting}[language=SQL,style=mystyle]
-- Q_case

SELECT e.PackageId
FROM Log l1, Log l2, EventData e
WHERE l1.TraceId=l2.TraceId AND l2.EventId=e.EventId AND
      l1.ActivityLabel='purchase package' AND
      l2.ActivityLabel='deliver package' AND
      l1.Lifecycle='complete' AND l2.Lifecycle='complete' AND
      e.Lifecycle='complete';


-- Q_viol

SELECT e.PackageId
FROM Log l1, Log l2, EventData e
WHERE l1.TraceId=l2.TraceId AND l2.EventId=e.EventId AND
      l1.ActivityLabel='purchase package' AND
      l2.ActivityLabel='deliver package' AND
      l1.Lifecycle='complete' AND l2.Lifecycle='complete' AND
      e.Lifecycle='complete' AND
      (
        DATE(l2.Timestamp) - DATE(l1.Timestamp) < 2
        OR
        DATE(l2.Timestamp) - DATE(l1.Timestamp) > 5
      );
\end{lstlisting}
\end{example}

\section{Compliance Monitoring as Incremental View Maintenance}
\label{sec:monitor}

\begin{figure}
\centering
\includegraphics[scale=0.28]{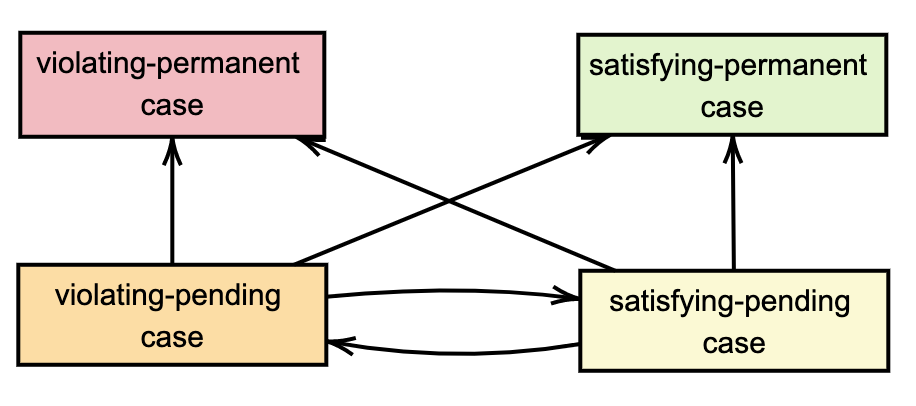}
\caption{\label{fig:states} A transition diagram of the different states that
a case could be in with respect to some constraint.  The diagram shows the
possible ways the state of a case can change as time progresses.  Not shown
in the diagram is that a case can also simply cease to be a case; furthermore,
new cases appear.}
\end{figure}

Now, if we want to monitor a constraint \emph{dynamically}, we will have to
refine our definition.  The reason is that the database instance representing
the execution log is continuously progressing.  Thus, the database instance
will contain the data of running (non-completed) process instances along with
the completed process instances.  Hence, at any moment, any case that is
subjected to some constraint will be in one of four different states \cite{BaLS11,MMWA11,DDGM14}:
\begin{inparaenum}[1)]
  \item a \emph{permanently} violating state;
  \item a \emph{permanently} satisfying state;
  \item a \emph{currently} violating state that may later be in a satisfying
  state as a result of the occurrences of new events; and
  \item similarly, a \emph{currently} satisfying state that may later be in a
  violating state.
\end{inparaenum} We will refer to the last two states as \emph{pending}
states.  Figure~\ref{fig:states} shows the different states and how a case
could change its state upon the occurrence of new events.  Notice that it depends on the constraint under study whether all such four states have to be actually considered, or whether instead
the constraint only requires a subset thereof.
Example~\ref{ex:simple} discusses a
simple constraint such that we can have its cases belonging to the different
states.

Regardless of the formal tools, languages, approaches, there is always a
``methodology" to go from informal specifications to formal realization.

\begin{figure}
\centering
\includegraphics[scale=0.32]{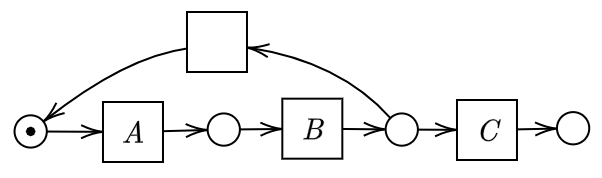}
\caption{\label{fig:pABC} A process model of an example process.}
\end{figure}

\begin{figure}
  \centering
  \includegraphics[scale=0.32]{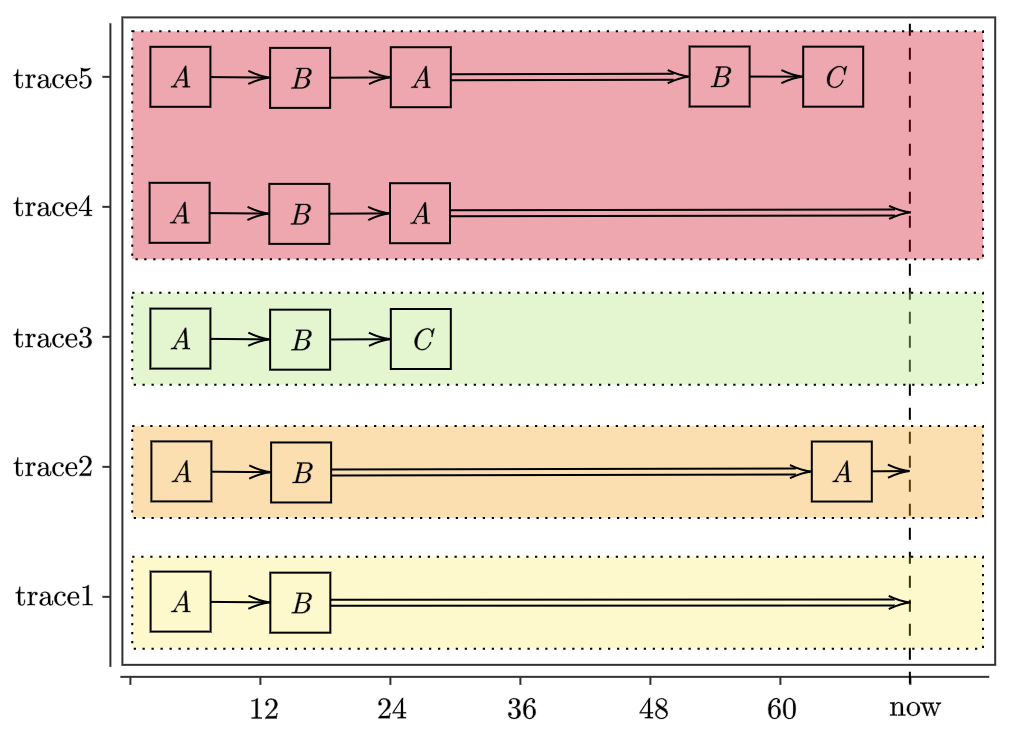}
  \caption{\label{fig:traces} The plot contains five different traces of
  the process whose model is shown in Figure~\ref{fig:pABC}.  The x-axis
  represents 12-hour intervals.  In a trace, double arrows ($\Rightarrow$)
  (respectively, single arrows ($\rightarrow$)) denote time intervals that
  are longer (respectively, equal or shorter) than 20 hours.  Each of the
  five traces is coloured based on its state at the ``now'' point with regard
  to the constraint \constraint{Every instance of activity $A$ must be directly
  followed by an instance of activity $B$ within 20 hours}.  For an example,
  the fifth trace is in a \emph{violating-permanent} state as the time span
  between the second execution of activities $A$ and $B$ is greater than the
  20-hour interval.}
\end{figure}

\begin{example}[Monitoring ``Followed-By" Constraint]\label{ex:simple}
  Consider a process that comprises three activities with the labels $A, B$,
  and $C$ whose process model is shown in Figure~\ref{fig:pABC}. Let the
  constraint that is imposed on this process be \constraint{Every instance of
  activity $A$ must be directly followed by an instance of activity $B$ within
  20 hours}.  In Figure~\ref{fig:traces}, we show five traces of that process
  which correspond to five cases as the constraint is an intra-instance one.
  The states of those five traces are distributed among the four different
  states.
\end{example}

\begin{definition}[Constraint, Compliance monitoring Variant]
\label{def:monitoring}
A \emph{constraint} $C$ is represented by four queries $(\qall, \qpermv,$
$\qpendv, \qpends)$, where $\qall$ returns all the cases subjected to the
constraint $C$, $\qpermv$ returns the \emph{permanently} violating cases,
$\qpendv$ returns the violating cases that later could be changed to
non-violating cases, while $\qpends$ returns the satisfying cases that later
could be changed to violating cases. (The cases not returned by none of these
three queries, are then the ones defined by $\qperms$.)
On any database instance, $\qpermv$, $\qpendv$, and $\qpends$ always return
three mutually exclusive subsets of $\qall$.
\end{definition}

\begin{remark}
  Typically the query $\qv$ in the post-mortem checking variant corresponds
  to the union of the pair $\qpermv$ and $\qpendv$ in the compliance monitoring
  variant.  Similarly, the query $\qs$ corresponds to the pair $\qperms$ and
  $\qpends$.
\end{remark}

\begin{example}[Monitoring Same Shipping Car Constraint]\label{ex:same_car2}
Consider the same constraint as in Example~\ref{ex:same_car1}.  The queries
representing this constraint can be defined as follows (where, $\qall$ and
$\qpermv$ are defined as $\qall$ and $\qv$ of Example~\ref{ex:same_car1};
respectively.):

\begin{lstlisting}[language=SQL,style=mystyle]
-- Q_sat-pending

SELECT e.CarId, DATE(l.Timestamp)
FROM Log l, EventData e
WHERE l.EventId=e.EventId AND
      l.ActivityLabel='deliver package' AND
      l.Lifecycle='complete' AND e.Lifecycle='complete' AND
      DATE(l.Timestamp)=CURRENT_DATE
GROUP BY e.CarId, DATE(l.Timestamp)
HAVING COUNT(e.PackageId) <= 7;
\end{lstlisting}
Note that $\qpendv$  will always be empty for this constraint.
\end{example}

\section{Experiments}
\label{sec:exp}

DBToaster is a state-of-the-art incremental query
processor~\cite{KennedyAK11,KochAKNNLS14,DBToasterWeb}.  As a proof-of-concept
of our approach, we tested DBToaster on some of the constraints from the work
of Winter et al.\ on automatic discovery of \isc~\cite{WinterSR20}.
Specifically, we worked with the constraints ISC1, ISC2a, ISC2b, ISC3, and ISC4
from the paper.  We have also used the execution logs provided by these authors
as sample input data~\cite{ExecutionLogData}.
To manage our experiments, we performed some preprocessing steps that are
mentioned in the Appendix~\ref{app:expPrep}.  To assess the feasibility and
usability of our approach, we have designed some experiments that ran over the
mentioned five constraints.  The results of these experiments are discussed
in Sections~\ref{sub:rt},~\ref{sub:qs}, and~\ref{sub:tc}.
At the beginning, we give a brief demonstration on the processes and the
constraints used in the experiments in Section~\ref{sub:expModel}.

\subsection{Experiment Data}\label{sub:expModel}

\begin{figure}
  \centering
  \includegraphics[width=\textwidth]{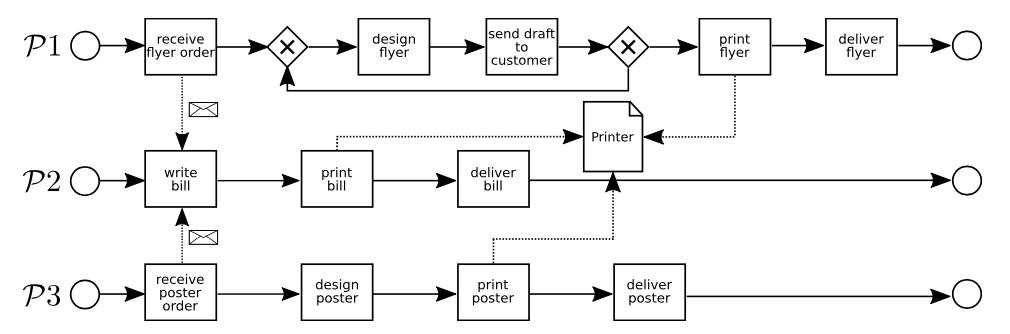}
  \caption{\label{fig:expProcesses} A figure showing the three process models
  whose executions are used in the experiments~\cite{WinterSR20}.}
\end{figure}

The constraints used in the experiments are expressed over the three processes
whose models are shown in Figure~\ref{fig:expProcesses}.  In the Figure, we
have ``Flyer Order", ``Poster Order", and ``Bill" processes that are labelled
as $P1$, $P3$, and $P2$, respectively.

The ``Flyer Order" process and ``Poster Order" process are quite similar.
Both processes begin by the activity of receiving the order.  This is
followed by designing the order activity, that is later followed by printing
the order.  In the end, the printed order is delivered.  The only difference
between the two processes is the extra activity of sending the design to the
customer for confirmation before the printing proceeds.  This is only part of
the ``Flyer Order" process.  The customer either accepts the design, then the
process proceeds as already mentioned.  Otherwise, if the customer rejects the
design then the order is redesigned and the same happens until the customer is
satisfied with the flyer design.  That explains the loop appearing in $P1$.
Any order whether it is for a flyer or a poster, has a corresponding initiated
``Bill" process.  This process is quite simple, it begins by the activity
of writing the bill, then the bill is printed and later delivered.  Moreover,
as you can see from the Figure, the printers are considered a shared resource
between all the processes.

The constraints used in the experiments are the following:
\begin{description}
   \item{ISC1} There is exactly one delivery activity per day in which all the
   finished orders/bills of that day so far are delivered to the post office
   simultaneously.

   \item{ISC2a} All print jobs must be completed within 10 minutes in at least
   95\% of all cases per month.

   \item{ISC2b} Printer 1 may only print 10 times per day.

   \item{ISC3} If a flyer or poster order is received $P2$ (i.e., bill process)
   is started afterwards.  Moreover, the corresponding bill process must be
   started before the order is delivered to the post office.

   \item{ISC4} Printing jobs that require different paper formats (i.e., A4
   and Poster formats) cannot be printed concurrently on one printer where
   concurrently means that one job starts, and before it finishes, the other
   starts.

\end{description}
We slightly modified the original constraints~\cite{WinterSR20} to better
match with the log data~\cite{ExecutionLogData}.

\subsection{Running Time}\label{sub:rt}

\begin{figure}
  \centering
  \includegraphics[width=\textwidth]{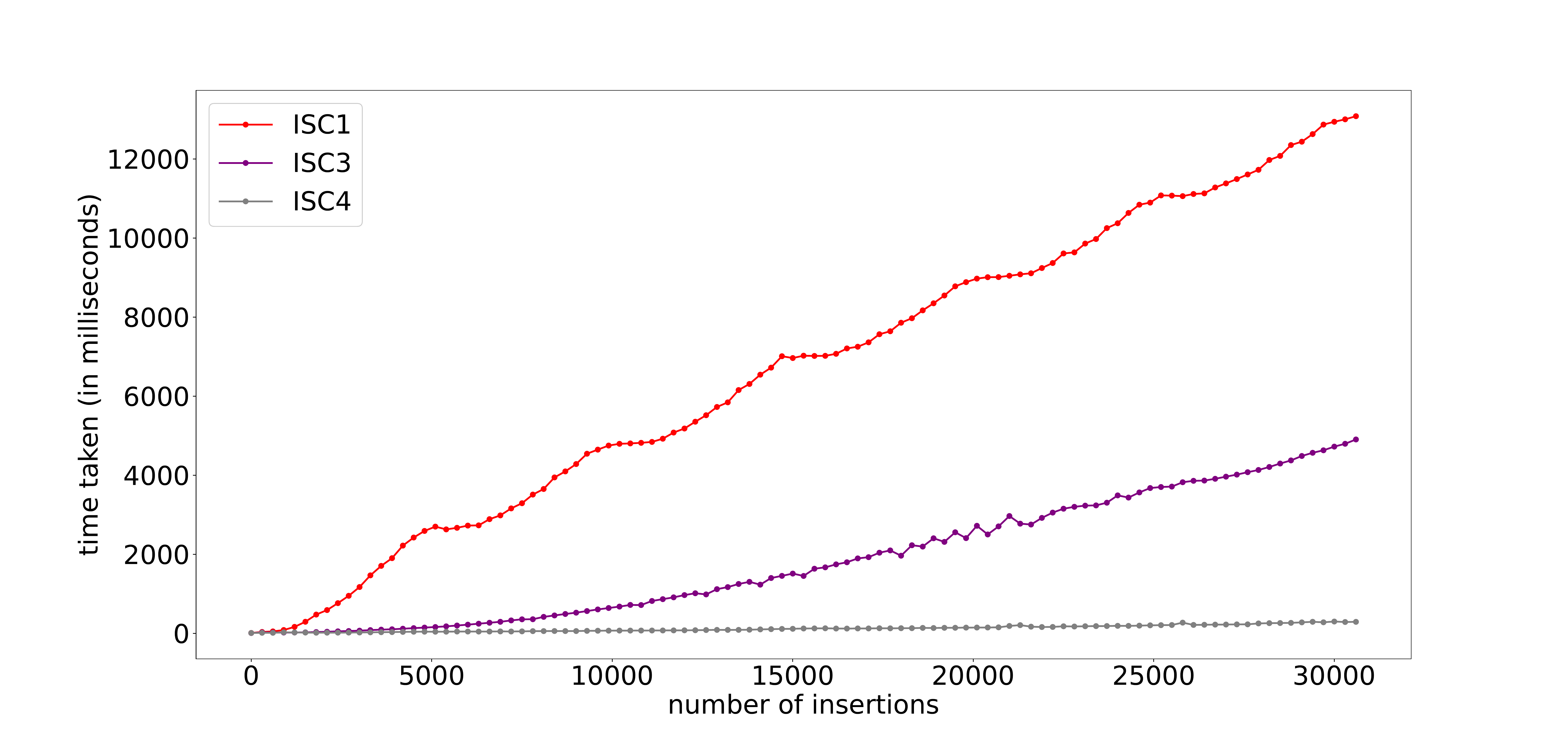}
  \caption{\label{fig:runTime} A plot of the running time (in milliseconds)
  taken to monitor the constraints ISC1, ISC3 and ISC4.  The running time of the
  constraints ISC2a and ISC2b are omitted since they are quite similar to ISC4.}
\end{figure}

The running time of three of the five monitored constraints is reported in
Figure~\ref{fig:runTime}, which shows averages over 10 runs.  The time is
reported for every 300 insertions with total insertions 30636 (the number of
events in the dataset).  This experiment was performed on a personal laptop
running macOS 12.2.1 with RAM of 16 GB and processor speed of 2.6Hz.

The slope of each curve is indicative of the average time needed, per event, to
maintain the queries defining the constraint.  We can see that this line is
significantly higher for the first constraint; indeed, this constraint requires
rather complex \sql~queries (shown in Appendix~\ref{app:expSQL}).
For tested constraints ISC1 and ISC3, the slopes of these lines are less than
half a millisecond, respectively less than 1/6th of a millisecond.  For tested
constraints ISC2a, ISCb and ISC4, the slopes are less than 1\% of a millisecond.

\subsection{Sizes of Queries}\label{sub:qs}

\begin{figure}
     \centering
     \begin{subfigure}[t]{0.49\textwidth}
         \centering
         \includegraphics[width=\textwidth]{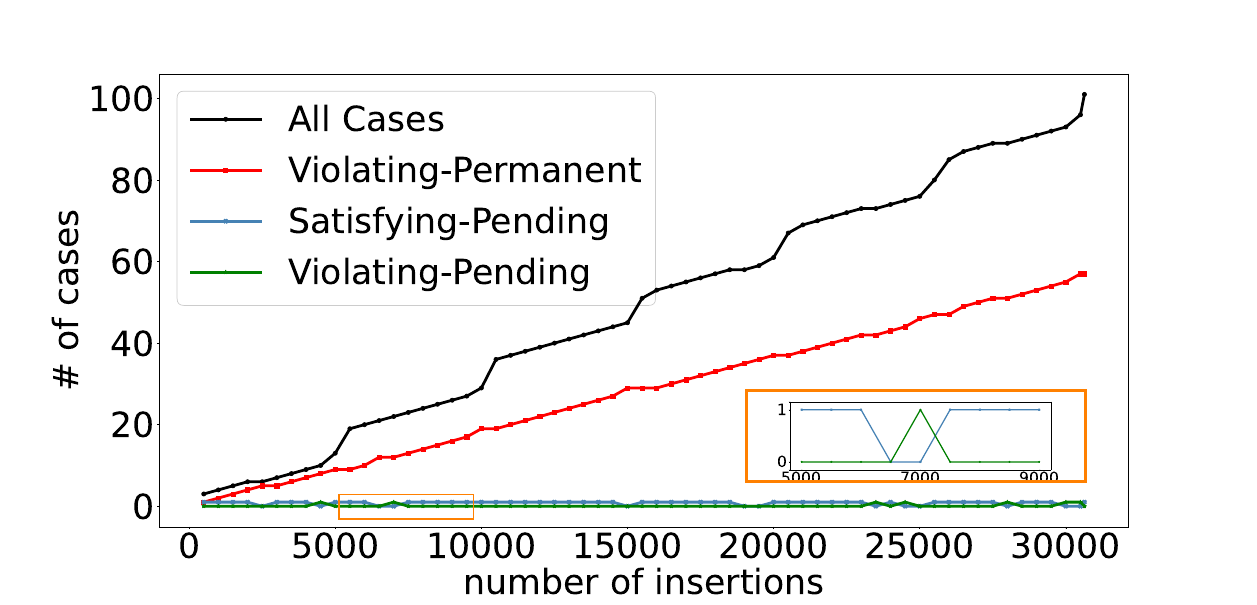}
         \caption{Plot of ISC1}
         \label{fig:ISC1QS}
     \end{subfigure}
     \hfill
     \begin{subfigure}[t]{0.49\textwidth}
         \centering
         \includegraphics[width=\textwidth]{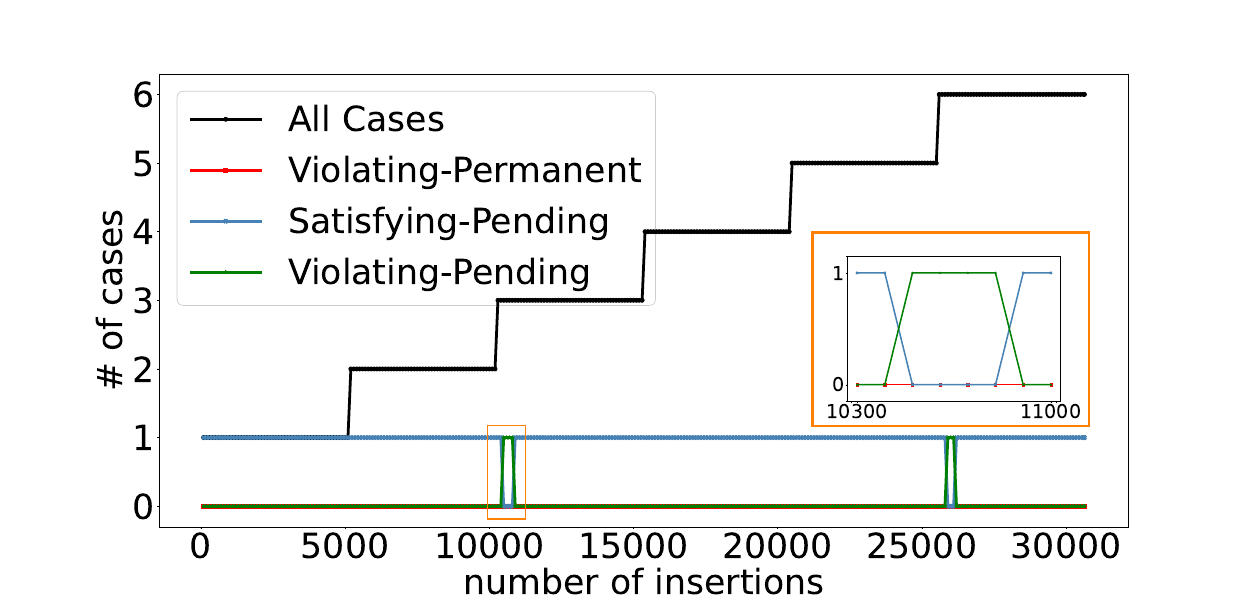}
         \caption{Plot of ISC2a}
         \label{fig:ISC2aQS}
     \end{subfigure}
     \qquad
     \begin{subfigure}[b]{0.49\textwidth}
         \centering
         \includegraphics[width=\textwidth]{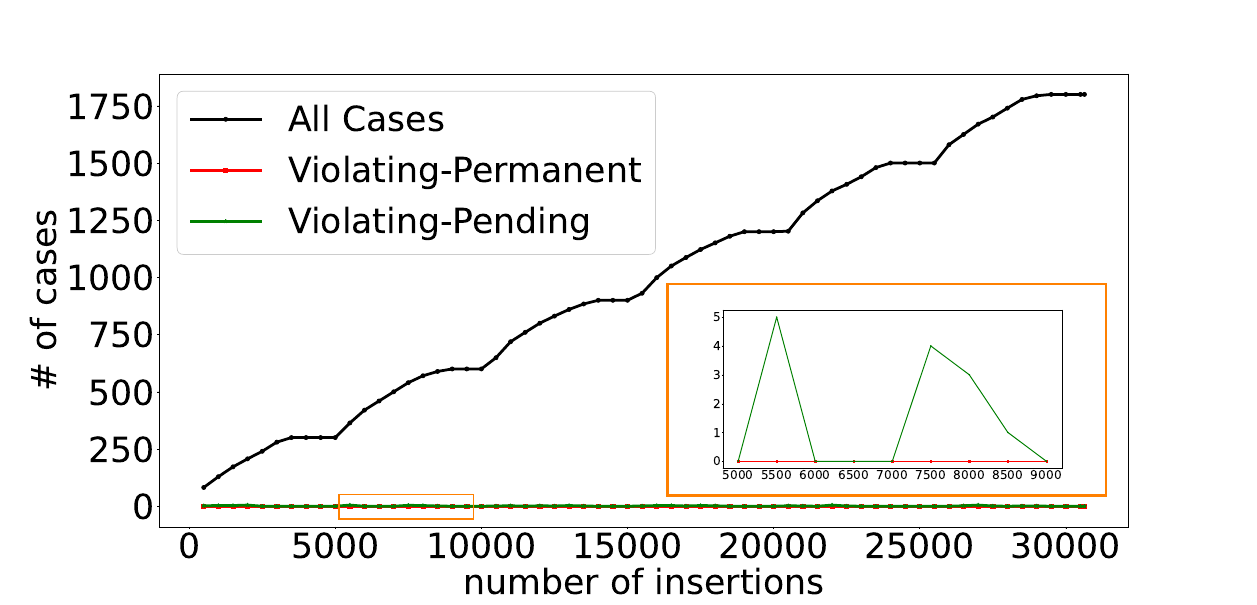}
         \caption{Plot of ISC3}
         \label{fig:ISC3QS}
     \end{subfigure}
     \hfill
     \begin{subfigure}[b]{0.49\textwidth}
         \centering
         \includegraphics[width=\textwidth]{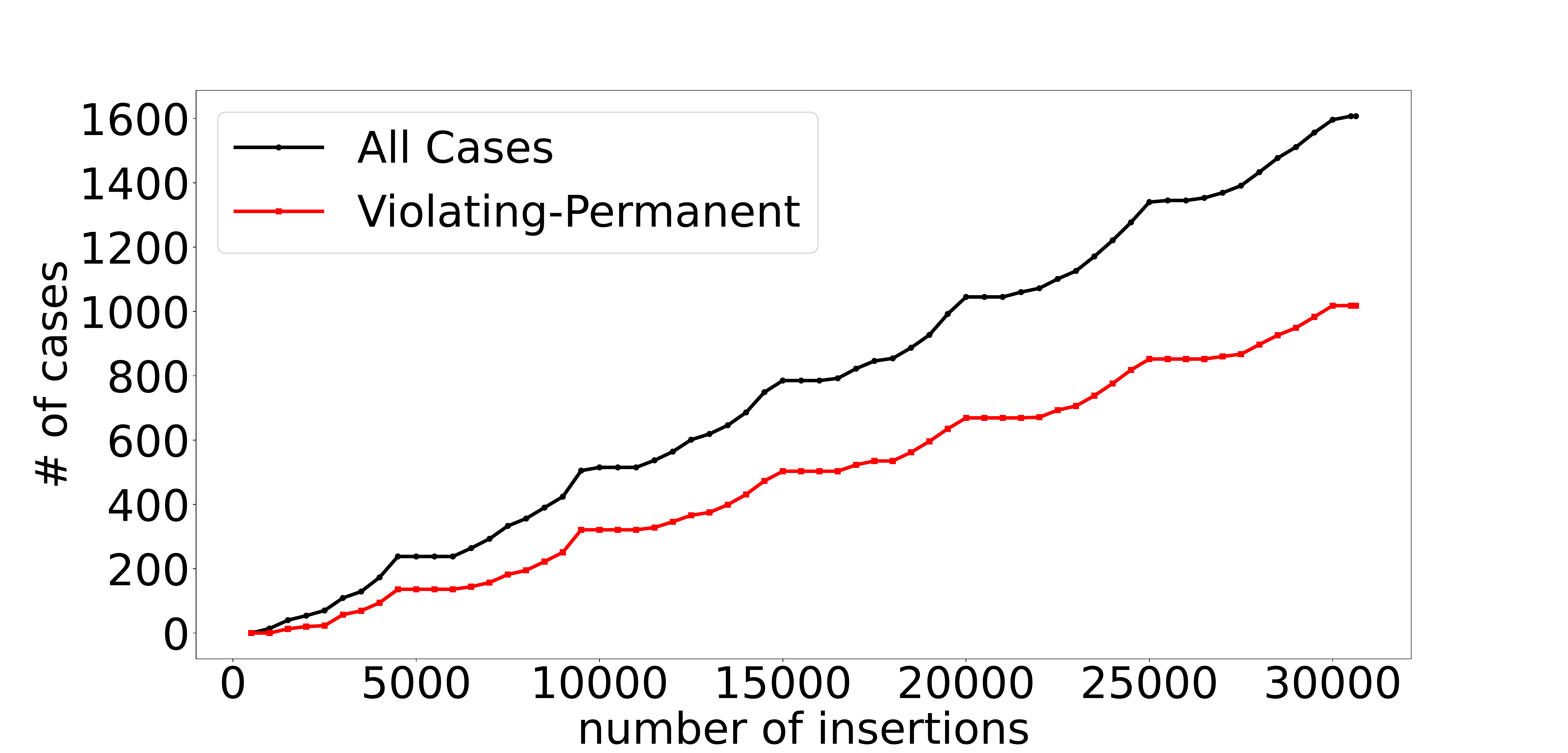}
         \caption{Plot of ISC4}
         \label{fig:ISC4QS}
     \end{subfigure}
    \caption{Plots of the size of each of the queries of the tested constraints.
    ISC2b is not shown as it has the same cases as ISC1 and has similar behavior
    to ISC3, which are shown.  Since our measurements consist of 600 data points
    (even 3000 for ISC2a), the plots are at rather high scale.  To show more
    detail, we provide insets that zoom in on selected regions (orange rectangles).}
    \label{fig:ISCQS}
\end{figure}

The size (i.e., the number of cases) of each of the queries defining four of
the five monitored constraints is reported and plotted relative to time (i.e.,
the number of insertions).  This can show us how the cases are changing
their status (pending or permanent, violating or satisfying).  A plot
for each of the four constraints is provided by Figure~\ref{fig:ISCQS}.
The query size is reported every 500 insertions except for ISC2a
which is done every 100 insertions instead, as it displays a more
fine-grained behavior.

\subsection{Tracing Cases}\label{sub:tc}
\begin{figure}
  \centering
  \includegraphics[width=\textwidth]{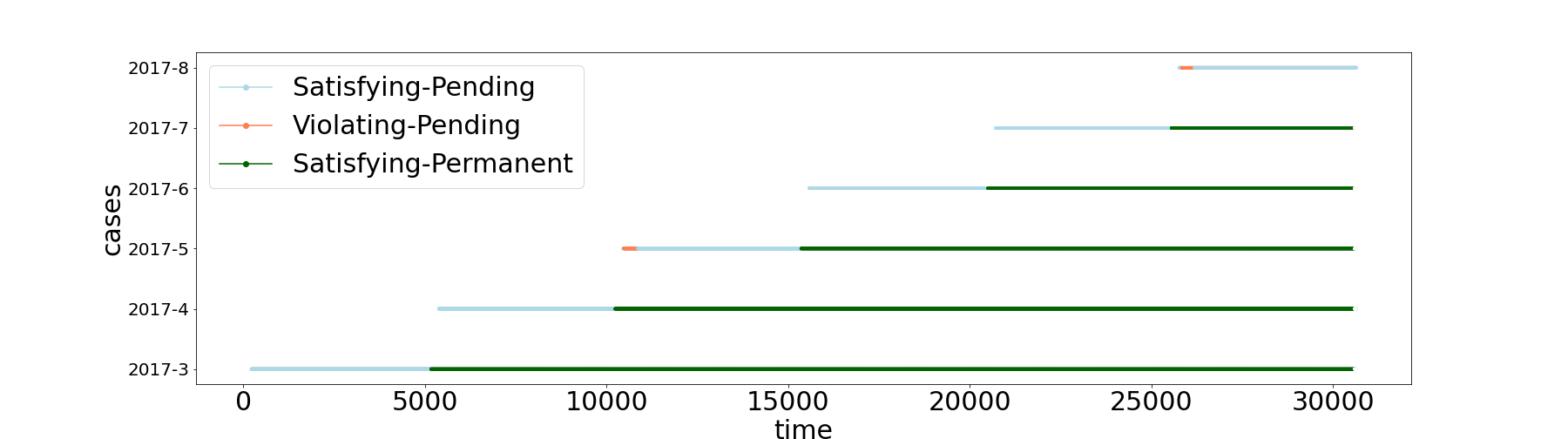} 
  \caption{\label{fig:c2SC} A plot showing the different cases of ISC2a and
  how each of the cases is changing its status through time.  From the
  previous plots, we see that in total we have six cases for this constraint.
  The cases according to this constraint are months.}
\end{figure}

\begin{figure}
  \centering
  \includegraphics[width=\textwidth]{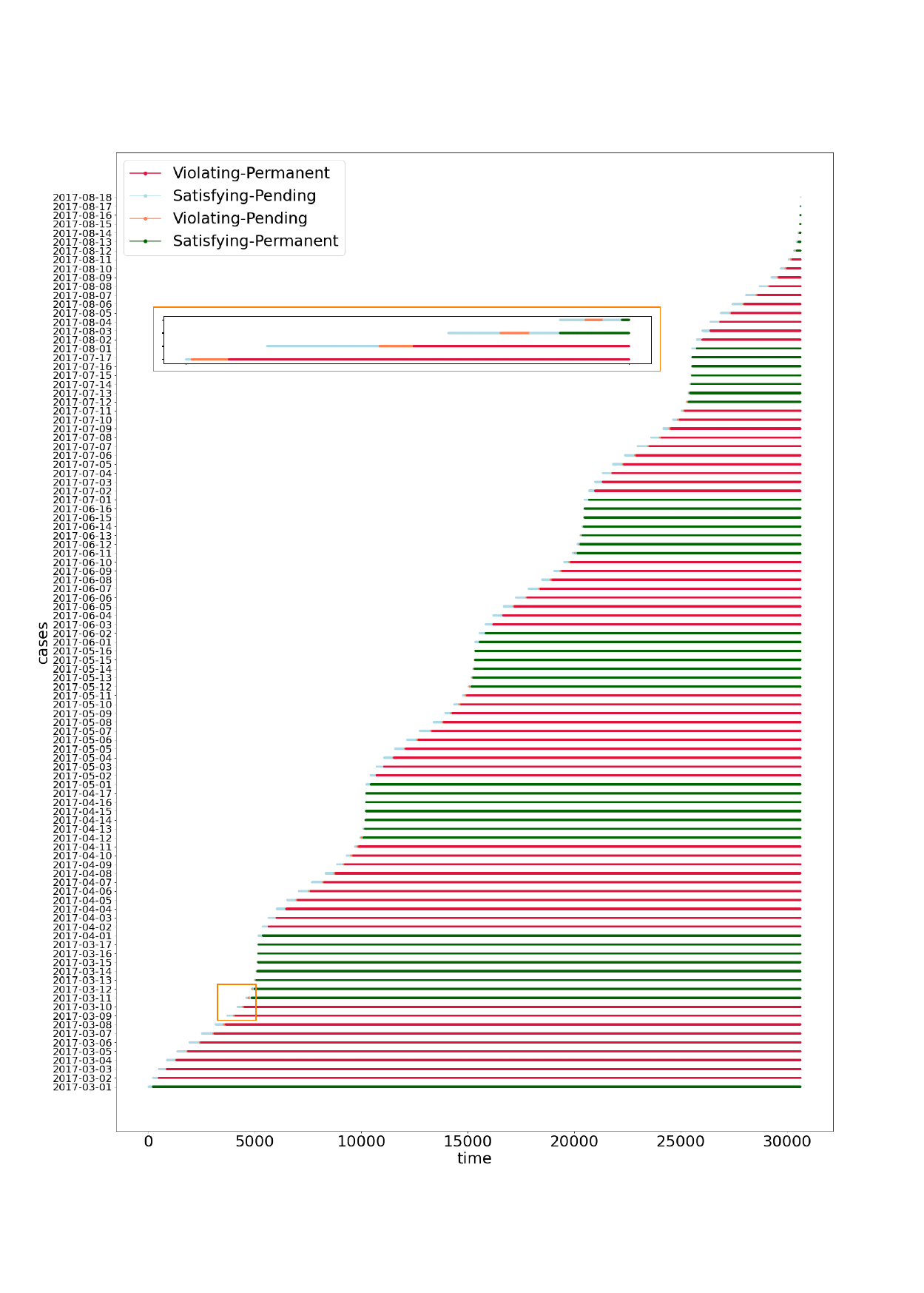} 
  \caption{\label{fig:c1SC} A plot showing the 101 different cases (days) of
  ISC1 and how each of the cases is changing its status through time.  Here the
  measurement consists of 30600 data points per case, so the plot is at a very
  high scale.  The inset shows more detail by zooming on the selected region
  (orange rectangle).}
\end{figure}

For ISC2a and ISC1, we show in Figures~\ref{fig:c2SC} and~\ref{fig:c1SC} the
evolution in status of all the individual cases over time.  This illustrates
that our approach is compatible with monitoring on a very detailed level.

\section{Sequence Data Extensions of Query Languages}
\label{sec:seqdlog}

We have mentioned before that any data model with a sufficiently expressive
query language can be used to express the constraints.  Although, we chose
to work with the relational data model with \sql~for the reasons we mentioned,
it is interesting to briefly discuss query languages for the relational data
model extended with sequences~\cite{AnglesABBFGLPPS18,ShenDNR07}.  Indeed, a
trace is a sequence of events.  Hence, representing the relative order of the
events is quite natural in a sequence data model.  This level of abstraction,
of viewing traces as sequences of abstract events, is often assumed when
working with temporal and dynamic
logics~\cite{GiacomoFMP21,PesicSA07,GiacomoMGMM14}.

Sequence Datalog~\cite{SeqDL,bonnermecca_sequences,meccabonner_termination} is
an extension of the query language Datalog, to work with sequences as first
class citizens.   We will briefly showcase this language by considering a
typical example of a constraint that is handled using the temporal logic.

\begin{example}[Strict Sequencing~\cite{GiacomoFMP21}]
  Let $\mathtt a$ and $\mathtt b$ be two activities.  Consider that we want to
  verify that the two activities are restricted by a \emph{strict sequencing}
  relation, which is one of the standard \emph{ordering}
  relations~\cite{vanderAalstWM2004}.  There is a strict sequencing relation
  between $\mathtt a$ and $\mathtt b$ if the log satisfies the following:
  \begin{itemize}
    \item there exists a trace where $\mathtt a$ is immediately followed by
    $\mathtt b$; and
    \item there are not any traces where $\mathtt b$ is immediately followed
    by $\mathtt a$.
  \end{itemize}
There are two possible violations of this constraint.  The first is not having
a trace with $\mathtt b$ directly following $\mathtt a$.  The other is having
a trace with $\mathtt a$ directly following $\mathtt b$.

For the purpose of expressing this constraint, assume we have the following
schema for the $\mathtt{Log}$ relation: $\mathtt{(TraceId,Events)}$, where
$\mathtt{Events}$ are just a sequence of labels of activities.  Then, this
constraint can be expressed by the following Sequence Datalog program.
{\small
\begin{lstlisting}
a_before_b():- Log(@traceId, $pre.a.b.$post).

violation():- +a_before_b().
violation():- Log(@traceId, $pre.b.a.$post).
\end{lstlisting}}
This program illustrates a number of Sequence Datalog features:
\begin{itemize}
  \item the dot is the concatenation operator.
  \item \verb|@traceId| is an \emph{atomic} variable (indicated by the
  \verb|@| symbol) representing atomic values (in this case, trace identifiers).
  \item \verb|$pre| and \verb|$post| are \emph{sequence} variables
  (indicated by the \verb|$| symbol) representing (possibly empty) sequences of
  atomic values.
\end{itemize}
\end{example}
The utility of using Sequence Datalog can be appreciated if we compare the
above program with the same query expressed in \sql.

\begin{lstlisting}[language=SQL,style=mystyle]
(
  SELECT 0
  FROM Log l1, Log l2
  WHERE l1.TraceId=l2.TraceId AND l1.Timestamp < l2.Timestamp AND
        l1.ActivityLabel='b' AND l2.ActivityLabel='a' AND
        NOT EXISTS
        (
          SELECT *
          FROM Log l3
          WHERE l1.TraceId=l3.TraceId AND
                l1.Timestamp < l3.Timestamp AND
                l3.Timestamp < l2.Timestamp
        )
)
UNION
(
  SELECT 0
  WHERE NOT EXISTS
  (
    SELECT 0
    FROM Log l1, Log l2
    WHERE l1.TraceId=l2.TraceId AND l1.Timestamp < l2.Timestamp AND
          l1.ActivityLabel='a' AND l2.ActivityLabel='b' AND
          NOT EXISTS
          (
            SELECT *
            FROM Log l3
            WHERE l1.TraceId=l3.TraceId AND
                  l1.Timestamp < l3.Timestamp AND
                  l3.Timestamp < l2.Timestamp
          )
  )
);
\end{lstlisting}

\section{Discussion}
\label{sec:conc}

In this paper, we have looked into the problems of post-mortem checking and
compliance monitoring of constraints over business processes.  Specifically,
we focused on \isc~as recently introduced in the process mining field, and
caught attention since it refers to complex constraints that span multiple
process instances.  Although there have been extensive works on inventorying
and categorizing \isc s~\cite{Rinderle-MaGFMI16,WinterSR20}, a crisp definition
of what is or is not an \isc, however, seems to be elusive.  Indeed, the notion
of constraint is so broad that we propose to \emph{define} any constraint as
two or four queries posed against the database instance that represents a
(partial) execution log.  This approach gives us huge flexibility, moreover,
we gain a lot from advances in database technology as demonstrated in the
Experiments Section.

In using the DBToaster system for our experiments, we faced a few technical
issues.  The main challenge was that the Scala version of DBToaster gets
stuck when retrieving snapshots over the course of the insertions.  To
overcome this issue, to perform our measurements of counting cases over time,
and how they evolve their constraint satisfaction status, we only retrieved
a snapshot after an initial sequence of insertions.  We then restart the
measurement for one batch of insertions longer.  Another limitation is that
\sql~is not yet fully supported, although complex queries can be expressed.
This required us to sometimes rewrite queries in equivalent form.  Finally,
some built-in functions (e.g., on strings or dates) are missing from the Scala
version.  Thus, those experiments should be seen more of a proof-of-concept of
the feasibility of our approach.

In this discussion, we briefly touch upon the main difference between our
approach and the main approach that is used to monitor \isc.  This approach
is based on the Event Calculus (EC)~\cite{LyMMRA15,MontaliMCMA13,ma2016}.
Most monitoring systems that are based on EC are implemented using Prolog.
Using EC to express a constraint seems to be very \emph{procedural} albeit
being defined in logical programming language.  For example, to monitor a
constraint such as ISC2b, in EC one would define a rule that increments a
counter every time a printing event occurs.  At the end, that counter value
should be at most 10 as per the constraint.  Since this is done in Prolog,
this will be asking the SAT solver if there exists an extension of the given
sequence of events satisfying the specification of this counting process.
A similar approach was followed in the paper by Montali et al.\
\cite{MontaliMCMA13} to monitor business (intra-instance) constraint with
the EC\@.  Events come in time, and Prolog rules that fire every
new time instant, are used to check various constraints dynamically.
However, these incremental rules are manually implemented.  On the
contrary, using an incremental query processor shifts the focus on what the
queries (or constraints) themselves are rather than what the rules are that
are responsible for this incremental maintenance.  Hence, our approach is
more declarative.

At the end of this discussion, we mention a few points for further research.
Since there are some algorithms that are used to discover \isc~from execution
logs~\cite{WinterSR20}, and these algorithms search for explicit patterns,
one could define a common language to report the results of those algorithms
and use those results to automatically write the \sql~queries monitoring each
of the reported constraints. Thus, the whole process could be automated.  Also,
one could try to rewrite the same queries differently and evaluate how the
different formulations affect the running times to incrementally maintain them.

\subsection*{Acknowledgments}
We thank Stefanie Rinderle-Ma and J\"{u}rgen Mangler for initial discussions.

%
%
%
\bibliographystyle{splncs04}
\bibliography{database.bib}

\begin{thebibliography}{10}
\providecommand{\url}[1]{\texttt{#1}}
\providecommand{\urlprefix}{URL }
\providecommand{\doi}[1]{https://doi.org/#1}

\bibitem{Aalst12}
van~der Aalst, W.M.P.: {Process Mining: Overview and Opportunities}. {ACM}
  Trans. Manag. Inf. Syst.  \textbf{3}(2),  7:1--7:17 (2012).
  \doi{10.1145/2229156.2229157}

\bibitem{vanderAalstWM2004}
van~der Aalst, W.M.P., Weijters, T., Maruster, L.: Workflow mining: Discovering
  process models from event logs. {IEEE} Trans. Knowl. Data Eng.
  \textbf{16}(9),  1128--1142 (2004). \doi{10.1109/TKDE.2004.47}

\bibitem{SeqDL}
Aamer, H., Hidders, J., Paredaens, J., Van~den Bussche, J.: {Expressiveness
  within Sequence Datalog}. In: Libkin, L., Pichler, R., Guagliardo, P. (eds.)
  PODS'21: Proceedings of the 40th {ACM} {SIGMOD-SIGACT-SIGAI} Symposium on
  Principles of Database Systems, Virtual Event, China, June 20-25, 2021. pp.
  70--81. {ACM} (2021). \doi{10.1145/3452021.3458327}

\bibitem{AnglesABBFGLPPS18}
Angles, R., Arenas, M., Barcel{\'{o}}, P., Boncz, P.A., Fletcher, G.H.L.,
  Guti{\'{e}}rrez, C., Lindaaker, T., Paradies, M., Plantikow, S., Sequeda,
  J.F., van Rest, O., Voigt, H.: {G-CORE: A Core for Future Graph Query
  Languages}. In: Das, G., Jermaine, C.M., Bernstein, P.A. (eds.) Proceedings
  of the 2018 International Conference on Management of Data, {SIGMOD}
  Conference 2018, Houston, TX, USA, June 10-15, 2018. pp. 1421--1432. {ACM}
  (2018). \doi{10.1145/3183713.3190654}

\bibitem{BaLS11}
Bauer, A., Leucker, M., Schallhart, C.: Runtime verification for {LTL} and
  {TLTL}. {ACM} Trans. Softw. Eng. Methodol.  \textbf{20}(4),  14:1--14:64
  (2011)

\bibitem{bonnermecca_sequences}
Bonner, A., Mecca, G.: {Sequences, Datalog, and Transducers}. J.~Comput. Syst.
  Sci.  \textbf{57},  234--259 (1998)

\bibitem{CDMP22}
Calvanese, D., De~Giacomo, G., Montali, M., Patrizi, F.: Verification and
  monitoring for first-order {LTL} with persistence-preserving quantification
  over finite and infinite traces. In: Proceedings of the 31st International
  Conference on Artificial Intelligence (IJCAI-ECAI 2022). AAAI Press (2022),
  to appear

\bibitem{CaMC19}
Cardoso, E., Montali, M., Calvanese, D.: Representing and querying norm states
  using temporal ontology-based data access. In: Proceedings of the 23rd {IEEE}
  International Enterprise Distributed Object Computing Conference ({EDOC}
  2019). pp. 122--131. {IEEE} (2019)

\bibitem{ChirkovaY12Book}
Chirkova, R., Yang, J.: {Materialized Views}. Foundations and
  Trends{\textregistered} in Databases  \textbf{4}(4),  295--405 (2012).
  \doi{10.1561/1900000020}

\bibitem{ExecutionLogData}
{CRISP project at Universit\"{a}t Wien}: {Execution Logs Webpage}.
  \url{http://gruppe.wst.univie.ac.at/projects/crisp/index.php?t=discovery},
  accessed: 2022-04-29

\bibitem{DDGM14}
De~Giacomo, G., De~Masellis, R., Grasso, M., Maggi, F.M., Montali, M.:
  Monitoring business metaconstraints based on {LTL} and {LDL} for finite
  traces. In: Sadiq, S.W., Soffer, P., V{\"{o}}lzer, H. (eds.) Proceedings of
  the 12th International Conference on Business Process Management ({BPM}
  2014). Lecture Notes in Computer Science, vol.~8659, pp. 1--17. Springer
  (2014)

\bibitem{DDMM22}
De~Giacomo, G., De~Masellis, R., Maggi, F.M., Montali, M.: Monitoring
  constraints and metaconstraints with temporal logics on finite traces. {ACM}
  Trans. Softw. Eng. Methodol.  (2022), to appear

\bibitem{Del09}
Demri, S., Lazic, R.: {LTL} with the freeze quantifier and register automata.
  ACM Trans.\ on Computational Logic  \textbf{10}(3) (2009)

\bibitem{FdhilaGRMI16}
Fdhila, W., Gall, M., Rinderle{-}Ma, S., Mangler, J., Indiono, C.:
  {Classification and Formalization of Instance-Spanning Constraints in
  Process-Driven Applications}. In: Rosa, M.L., Loos, P., Pastor, O. (eds.)
  Business Process Management - 14th International Conference, {BPM} 2016, Rio
  de Janeiro, Brazil, September 18-22, 2016. Proceedings. Lecture Notes in
  Computer Science, vol.~9850, pp. 348--364. Springer (2016).
  \doi{10.1007/978-3-319-45348-4\_20}

\bibitem{PM4PyLink}
{Fraunhofer Institute for Applied Information Technology (FIT)}: {PM4Py
  Website}. \url{https://pm4py.fit.fraunhofer.de/}, accessed: 2022-04-29

\bibitem{GiacomoFMP21}
Giacomo, G.D., Felli, P., Montali, M., Perelli, G.: {HyperLDLf: a Logic for
  Checking Properties of Finite Traces Process Logs}. In: Zhou, Z. (ed.)
  Proceedings of the Thirtieth International Joint Conference on Artificial
  Intelligence, {IJCAI} 2021, Virtual Event / Montreal, Canada, 19-27 August
  2021. pp. 1859--1865. ijcai.org (2021). \doi{10.24963/ijcai.2021/256}

\bibitem{GiacomoMGMM14}
Giacomo, G.D., Masellis, R.D., Grasso, M., Maggi, F.M., Montali, M.:
  {Monitoring Business Metaconstraints Based on {LTL} and {LDL} for Finite
  Traces}. In: Sadiq, S.W., Soffer, P., V{\"{o}}lzer, H. (eds.) Business
  Process Management - 12th International Conference, {BPM} 2014, Haifa,
  Israel, September 7-11, 2014. Proceedings. Lecture Notes in Computer Science,
  vol.~8659, pp. 1--17. Springer (2014). \doi{10.1007/978-3-319-10172-9\_1}

\bibitem{GuptaM99Book}
Gupta, A., Mumick, I.S. (eds.): {Materialized Views: Techniques,
  Implementations, and Applications}. MIT Press, Cambridge, MA, USA (1999)

\bibitem{GuptaM95}
Gupta, A., Mumick, I.S.: {Maintenance of Materialized Views: Problems,
  Techniques, and Applications}. {IEEE} Data Eng. Bull.  \textbf{18}(2),  3--18
  (1995), \url{http://sites.computer.org/debull/95JUN-CD.pdf}

\bibitem{GuptaMS93}
Gupta, A., Mumick, I.S., Subrahmanian, V.S.: {Maintaining Views Incrementally}.
  In: Buneman, P., Jajodia, S. (eds.) Proceedings of the 1993 {ACM} {SIGMOD}
  International Conference on Management of Data, Washington, DC, USA, May
  26-28, 1993. pp. 157--166. {ACM} Press (1993). \doi{10.1145/170035.170066}

\bibitem{XESLink}
{IEEE XES Group}: {IEEE 1849-2016 XES Standard}.
  \url{https://www.xes-standard.org/}, accessed: 2022-05-11

\bibitem{ma2016}
Indiono, C., Mangler, J., Fdhila, W., Rinderle{-}Ma, S.: {Rule-Based Runtime
  Monitoring of Instance-Spanning Constraints in Process-Aware Information
  Systems}. In: Debruyne, C., Panetto, H., Meersman, R., Dillon, T.S., eva
  K{\"{u}}hn, O'Sullivan, D., Ardagna, C.A. (eds.) On the Move to Meaningful
  Internet Systems: {OTM} 2016 Conferences - Confederated International
  Conferences: CoopIS, C{\&}TC, and {ODBASE} 2016, Rhodes, Greece, October
  24-28, 2016, Proceedings. Lecture Notes in Computer Science, vol. 10033, pp.
  381--399 (2016). \doi{10.1007/978-3-319-48472-3\_22}

\bibitem{KennedyAK11}
Kennedy, O., Ahmad, Y., Koch, C.: {DBToaster: Agile Views for a Dynamic Data
  Management System}. In: Fifth Biennial Conference on Innovative Data Systems
  Research, {CIDR} 2011, Asilomar, CA, USA, January 9-12, 2011, Online
  Proceedings. pp. 284--295. www.cidrdb.org (2011),
  \url{http://cidrdb.org/cidr2011/Papers/CIDR11\_Paper38.pdf}

\bibitem{KochAKNNLS14}
Koch, C., Ahmad, Y., Kennedy, O., Nikolic, M., N{\"{o}}tzli, A., Lupei, D.,
  Shaikhha, A.: {DBToaster: Higher-order Delta Processing for Dynamic,
  Frequently Fresh Views}. {VLDB} J.  \textbf{23}(2),  253--278 (2014).
  \doi{10.1007/s00778-013-0348-4}

\bibitem{LyMMRA15}
Ly, L.T., Maggi, F.M., Montali, M., Rinderle{-}Ma, S., van~der Aalst, W.M.P.:
  {Compliance Monitoring in Business Processes: Functionalities, Application,
  and Tool-Support}. Inf. Syst.  \textbf{54},  209--234 (2015).
  \doi{10.1016/j.is.2015.02.007}

\bibitem{MMWA11}
Maggi, F.M., Montali, M., Westergaard, M., van~der Aalst, W.M.P.: Monitoring
  business constraints with linear temporal logic: An approach based on colored
  automata. In: Rinderle{-}Ma, S., Toumani, F., Wolf, K. (eds.) Proceedings of
  the 9th International Conference on Business Process Management ({BPM} 2011).
  Lecture Notes in Computer Science, vol.~6896, pp. 132--147. Springer (2011)

\bibitem{meccabonner_termination}
Mecca, G., Bonner, A.: {Query Languages for Sequence Databases: Termination and
  Complexity}. IEEE Transactions on Knowledge and Data Engineering
  \textbf{13}(3),  519--525 (2001)

\bibitem{MontaliMCMA13}
Montali, M., Maggi, F.M., Chesani, F., Mello, P., van~der Aalst, W.M.P.:
  {Monitoring Business Constraints with the Event Calculus}. {ACM} Trans.
  Intell. Syst. Technol.  \textbf{5}(1),  17:1--17:30 (2013).
  \doi{10.1145/2542182.2542199}

\bibitem{MurillasRA19}
de~Murillas, E.G.L., Reijers, H.A., van~der Aalst, W.M.P.: {Connecting
  databases with process mining: a meta model and toolset}. Softw. Syst. Model.
   \textbf{18}(2),  1209--1247 (2019). \doi{10.1007/s10270-018-0664-7}

\bibitem{PesicSA07}
Pesic, M., Schonenberg, H., van~der Aalst, W.M.P.: {DECLARE: Full Support for
  Loosely-Structured Processes}. In: 11th {IEEE} International Enterprise
  Distributed Object Computing Conference {(EDOC} 2007), 15-19 October 2007,
  Annapolis, Maryland, {USA}. pp. 287--300. {IEEE} Computer Society (2007).
  \doi{10.1109/EDOC.2007.14}

\bibitem{Rinderle-MaGFMI16}
Rinderle{-}Ma, S., Gall, M., Fdhila, W., Mangler, J., Indiono, C.: {Collecting
  Examples for Instance-Spanning Constraints}. arXiv:1603.01523 (2018)

\bibitem{ShenDNR07}
Shen, W., Doan, A., Naughton, J.F., Ramakrishnan, R.: {Declarative Information
  Extraction Using Datalog with Embedded Extraction Predicates}. In: Koch, C.,
  Gehrke, J., Garofalakis, M.N., Srivastava, D., Aberer, K., Deshpande, A.,
  Florescu, D., Chan, C.Y., Ganti, V., Kanne, C., Klas, W., Neuhold, E.J.
  (eds.) Proceedings of the 33rd International Conference on Very Large Data
  Bases, University of Vienna, Austria, September 23-27, 2007. pp. 1033--1044.
  {ACM} (2007)

\bibitem{DBToasterWeb}
{The DBToaster Consortium}: {DBToaster Webpage}.
  \url{https://dbtoaster.github.io/index.html}, accessed: 2022-04-29

\bibitem{WarnerA06}
Warner, J., Atluri, V.: {Inter-instance Authorization Constraints for Secure
  Workflow Management}. In: Ferraiolo, D.F., Ray, I. (eds.) 11th {ACM}
  Symposium on Access Control Models and Technologies, {SACMAT} 2006, Lake
  Tahoe, California, USA, June 7-9, 2006, Proceedings. pp. 190--199. {ACM}
  (2006). \doi{10.1145/1133058.1133085}

\bibitem{WinterSR20}
Winter, K., Stertz, F., Rinderle{-}Ma, S.: {Discovering Instance and Process
  Spanning Constraints from Process Execution Logs}. Inf. Syst.  \textbf{89},
  101484 (2020). \doi{10.1016/j.is.2019.101484}

\end{thebibliography}

\appendix

\section{Experiments Preprocessing}\label{app:expPrep}

As mentioned in the main paper, we performed the following preprocessing
steps to manage our experiments:
\begin{enumerate}
\item The execution logs are given in \xes~format; we converted them to
\csv~using the process mining python library PM4Py~\cite{PM4PyLink}.
\item The events from the different processes are merged and sorted based
on the timestamp attribute.  In this way, we simulate a stream of events
suitable for monitoring.
\item For each of the selected constraints, we formulated appropriate
\sql~queries defining the cases, the violations, the pending violations,
and the pending satisfying cases, following our methodology described in
Definition~\ref{def:monitoring}.
\item DBToaster takes these queries and produces an executable program
(\jar~file) that allows to communicate with the queries while being
incrementally maintained.
\item Lastly, we have implemented a Scala program for each of the
constraints that reads the \csv~file and communicates with the incremental
processor from Step 4 by sending the events as insertions and asking for
the intermediate results of the queries.
\end{enumerate}

\section{SQL Queries of Constraints}\label{app:expSQL}

We begin this section by Table~\ref{tab:sqlFeatures} that summarizes the
\sql~features used in the monitored constraints.  In our queries, we use
a view \texttt{Events} which is defined by the natural join of \texttt{Log}
and \texttt{EventData}.  Afterwards, we show these \sql~queries.
\begin{table}
\caption{A table showing the \sql~features used in the queries expressing
each of the monitored five constraints.}\label{tab:sqlFeatures}
\centering
\setlength{\tabcolsep}{6pt}
\resizebox{0.95\textwidth}{!}{%
\begin{tabular}{|l|c|c|c|c|c|}
\hline
ISC & Aggregation & OR & Existence Check & Negation & Double Negation \\
\hline
ISC1 & no & no & yes & yes & yes \\ \hline
ISC2a & yes & no & yes & yes & no \\ \hline
ISC2b & yes & no & no & yes & no \\ \hline
ISC3 & no & yes & yes & yes & no \\ \hline
ISC4 & no & no & no & yes & no \\ \hline
\end{tabular}}
\end{table}

\subsection{Queries of ISC1}
\begin{lstlisting}[language=SQL,style=mystyle]
-- Q_case
SELECT DISTINCT(DATE(Timestamp)) FROM Events;

-- Q_viol
SELECT DATE(e.Timestamp)
FROM Events e, Events e4
WHERE e.Lifecycle = 'start' AND e.ActivityLabel LIKE 'deliver%'
      AND DATE(e.Timestamp) = DATE(e4.Timestamp)
      AND e.Timestamp < e4.Timestamp
      AND ( EXISTS
        ( SELECT *
          FROM Events e2
          WHERE DATE(e.Timestamp) = DATE(e2.Timestamp) 
                AND e2.Lifecycle = 'complete'
                AND e2.ActivityLabel LIKE 'print%'
                AND e2.Timestamp < e.Timestamp 
                AND ( NOT EXISTS
                    ( SELECT *
                      FROM Events e3
                      WHERE e3.ProcessId = e2.ProcessId 
                            AND e3.TraceId = e2.TraceId
                            AND e3.ActivityLabel LIKE 'deliver%'
                            AND e3.Lifecycle = 'start'
                            AND e3.Timestamp = e.Timestamp
                    ) ) ) );

-- Q_sat-pending
SELECT DATE(c.Timestamp)
FROM CURR_DAY c
WHERE DATE(c.Timestamp) NOT IN
( SELECT DATE(e.Timestamp)
  FROM Events e
  WHERE e.Lifecycle = 'start' AND e.ActivityLabel LIKE 'deliver%'
        AND ( ( EXISTS
            ( SELECT *
              FROM Events e2
              WHERE DATE(e.Timestamp) = DATE(e2.Timestamp) 
                    AND e2.Lifecycle = 'complete'
                    AND e2.ActivityLabel LIKE 'print%' 
                    AND e2.Timestamp < e.Timestamp
                    AND ( NOT EXISTS 
                        ( SELECT *
                          FROM Events e3
                          WHERE e3.ProcessId = e2.ProcessId 
                                AND e3.TraceId = e2.TraceId
                                AND e3.ActivityLabel LIKE 'deliver%' 
                                AND e3.Lifecycle = 'start'
                                AND e3.Timestamp = e.Timestamp
                        ) ) ) ) ) );

-- Q_viol-pending
SELECT DATE(c.Timestamp)
FROM CURR_DAY c, Events e
WHERE e.Lifecycle = 'start' AND e.ActivityLabel LIKE 'deliver%'
      AND DATE(e.Timestamp) = DATE(c.Timestamp)
      AND ( EXISTS
        ( SELECT *
          FROM Events e2
          WHERE DATE(e.Timestamp) = DATE(e2.Timestamp) 
                AND e2.Lifecycle = 'complete'
                AND e2.ActivityLabel LIKE 'print%' 
                AND e2.Timestamp < e.Timestamp
                AND ( NOT EXISTS
                  ( SELECT *
                    FROM Events e3
                    WHERE e3.ProcessId = e2.ProcessId 
                          AND e3.TraceId = e2.TraceId
                          AND e3.ActivityLabel LIKE 'deliver%' 
                          AND e3.Lifecycle = 'start'
                          AND e3.Timestamp = e.Timestamp
                  ) ) ) )
      AND ( NOT EXISTS
        ( SELECT * 
          FROM Events e4
          WHERE DATE(e.Timestamp) = DATE(e4.Timestamp)
                AND e.Timestamp < e4.Timestamp
        ) );
\end{lstlisting}

\subsection{Queries of ISC2a}
In these queries, we use the terminology of $\tt e.Year$, $\tt e.Month$, $\tt e.Hour$,
and $\tt e.Minute$ for some $\tt Event \, e$.  You can consider these as shortcuts for
the following clauses, receptively:
\begin{itemize}
    \item $\tt EXTRACT(Year \, FROM \, e.Timestamp)$,
    \item $\tt EXTRACT(Month \, FROM \, e.Timestamp)$,
    \item $\tt EXTRACT(Hour \, FROM \, e.Timestamp)$, and
    \item $\tt EXTRACT(Minute \, FROM \, e.Timestamp)$.
\end{itemize}
\begin{lstlisting}[language=SQL,style=mystyle]
-- Q_case
SELECT m.Year, m.Month
FROM (
  SELECT e1.Year, e1.Month, COUNT(*) AS NumberOfMonthCases,
         SUM( CASE WHEN ((((e2.Hour - e1.Hour) * 60) + 
                           (e2.Minute - e1.Minute)) > 10)
              THEN 0 ELSE 1 END) AS NumberOfMonthFastCases
  FROM Events e1, Events e2
  WHERE e1.ProcessId = e2.ProcessId AND e1.TraceId = e2.TraceId AND
        e1.Lifecycle = 'start' AND e2.Lifecycle = 'complete' AND
        e1.ActivityLabel LIKE 'print%' AND e2.ActivityLabel LIKE 'print%'
  GROUP BY e1.Year, e1.Month
) as m;

-- Q_viol
SELECT m.Year, m.Month
FROM (
  SELECT e1.Year, e1.Month, COUNT(*) AS NumberOfMonthCases,
         SUM( CASE WHEN ((((e2.Hour - e1.Hour) * 60) + 
                           (e2.Minute - e1.Minute)) > 10)
              THEN 0 ELSE 1 END) AS NumberOfMonthFastCases
  FROM Events e1, Events e2
  WHERE e1.ProcessId = e2.ProcessId AND e1.TraceId = e2.TraceId AND
        e1.Lifecycle = 'start' AND e2.Lifecycle = 'complete' AND
        e1.ActivityLabel LIKE 'print%' AND e2.ActivityLabel LIKE 'print%'
  GROUP BY e1.Year, e1.Month
) as m
WHERE (m.NumberOfMonthFastCases*100/m.NumberOfMonthCases) <= 95 
      AND NOT EXISTS 
          ( SELECT * FROM CURR_MONTH c 
            WHERE c.Year = m.YEAR AND c.Month = m.Month );

-- Q_sat-pending
SELECT m.Year, m.Month
FROM (
  SELECT e1.Year, e1.Month, COUNT(*) AS NumberOfMonthCases,
         SUM( CASE WHEN ((((e2.Hour - e1.Hour) * 60) + 
                           (e2.Minute - e1.Minute)) > 10)
              THEN 0 ELSE 1 END) AS NumberOfMonthFastCases
  FROM Events e1, Events e2
  WHERE e1.ProcessId = e2.ProcessId AND e1.TraceId = e2.TraceId AND
        e1.Lifecycle = 'start' AND e2.Lifecycle = 'complete' AND
        e1.ActivityLabel LIKE 'print%' AND e2.ActivityLabel LIKE 'print%'
  GROUP BY e1.Year, e1.Month
) as m
WHERE (m.NumberOfMonthFastCases*100/m.NumberOfMonthCases) > 95 
      AND EXISTS 
        ( SELECT * FROM CURR_MONTH c 
          WHERE c.Year = m.Year AND c.Month = m.Month );

-- Q_viol-pending
SELECT m.Year, m.Month
FROM (
  SELECT e1.Year, e1.Month, COUNT(*) AS NumberOfMonthCases,
         SUM( CASE WHEN ((((e2.Hour - e1.Hour) * 60) + 
                           (e2.Minute - e1.Minute)) > 10)
              THEN 0 ELSE 1 END) AS NumberOfMonthFastCases
  FROM Events e1, Events e2
  WHERE e1.ProcessId = e2.ProcessId AND e1.TraceId = e2.TraceId AND
        e1.Lifecycle = 'start' AND e2.Lifecycle = 'complete' AND
        e1.ActivityLabel LIKE 'print%' AND e2.ActivityLabel LIKE 'print%'
  GROUP BY e1.Year, e1.Month
) as m
WHERE (m.NumberOfMonthFastCases*100/m.NumberOfMonthCases) <= 95 
      AND EXISTS 
        ( SELECT * FROM CURR_MONTH c 
          WHERE c.Year = m.Year AND c.Month = m.Month );
\end{lstlisting}

\subsection{Queries of ISC2b}
\begin{lstlisting}[language=SQL,style=mystyle]
-- Q_case
SELECT DISTINCT(DATE(Timestamp)) FROM Events;

-- Q_viol
SELECT DATE(t.Timestamp)
FROM (
  SELECT DATE(e.Timestamp), COUNT(e.TraceId) AS Uses
  FROM Events e
  WHERE e.Lifecycle = 'complete' AND e.ActivityLabel LIKE 'print%'
        AND e.Resource = 'printer1'
  GROUP BY DATE(e.Timestamp)
) AS t
WHERE t.Uses > 10;

-- Q_sat-pending
SELECT DATE(c.Timestamp)
FROM CURR_DAY c
WHERE c.Date NOT IN (
  SELECT DATE(t.Timestamp)
  FROM (
    SELECT DATE(e.Timestamp), COUNT(e.TraceId) AS Uses
    FROM Events e
    WHERE e.Lifecycle = 'complete' AND e.ActivityLabel LIKE 'print%'
          AND e.Resource = 'printer1'
    GROUP BY DATE(e.Timestamp)
  ) AS t
  WHERE t.Uses > 10
);
\end{lstlisting}

\subsection{Queries of ISC3}
\begin{lstlisting}[language=SQL,style=mystyle]
-- Q_case
SELECT ProcessId, TraceId FROM Events
WHERE ActivityLabel LIKE 'receive %';

-- Q_viol
SELECT e1.ProcessId, e1.TraceId
FROM Events e1
WHERE e1.ActivityLabel LIKE 'receive %' AND e1.Lifecycle = 'start' AND
(
  (
    ( EXISTS ( SELECT * FROM Events e2
               WHERE e1.ProcessId = e2.ProcessId 
                     AND e1.TraceId = e2.TraceId
                     AND e2.ActivityLabel LIKE 'deliver %'
                     AND e2.Lifecycle = 'complete' ) ) 
    AND
    ( NOT EXISTS ( SELECT * FROM Events e3
                   WHERE e3.ActivityLabel LIKE 'write bill'
                         AND e3.CustomerId = e1.CustomerId ) )
  ) -- P2 is not yet started but the order is delivered
  OR
  ( EXISTS ( SELECT * FROM Events e4
             WHERE e4.ActivityLabel LIKE 'write bill'
                   AND e4.CustomerId = e1.CustomerId
                   AND e4.Timestamp < e1.Timestamp )
  ) -- P2 is started before the order being received
);

-- Q_viol-pending
SELECT e1.ProcessId, e1.TraceId
FROM Events e1
WHERE e1.ActivityLabel LIKE 'receive %' AND e1.Lifecycle = 'start' 
      AND
        ( NOT EXISTS ( 
            SELECT * FROM Events e2
            WHERE e1.ProcessId = e2.ProcessId AND e1.TraceId = e2.TraceId
                  AND e2.ActivityLabel LIKE 'deliver %'
                  AND e2.Lifecycle = 'complete' ) )
      AND
        ( NOT EXISTS (
            SELECT * FROM Events e3
            WHERE e3.ActivityLabel LIKE 'write bill'
                  AND e3.CustomerId = e1.CustomerId )
        );
\end{lstlisting}

\subsection{Queries of ISC4}
\begin{lstlisting}[language=SQL,style=mystyle]
-- Q_case
SELECT e1.ProcessId, e1.TraceId, e3.ProcessId, e3.TraceId
FROM Events e1, Events e3
WHERE e1.ActivityLabel LIKE 'print%' AND e3.ActivityLabel LIKE 'print%'
      AND e1.Lifecycle = 'start'AND e3.Lifecycle = 'start'
      AND DATE(e1.Timestamp) = DATE(e3.Timestamp) 
      AND e1.Timestamp < e3.Timestamp
      AND ( 
        NOT EXISTS (
            SELECT * FROM Events e2
            WHERE e1.ProcessId = e2.ProcessId AND e1.TraceId = e2.TraceId
                  AND e2.ActivityLabel LIKE 'print%' 
                  AND e2.Lifecycle = 'complete' 
                  AND e1.Timestamp < e2.Timestamp 
                  AND e2.Timestamp < e3.Timestamp
        ) );

-- Q_viol
SELECT e1.ProcessId, e1.TraceId, e3.ProcessId, e3.TraceId
FROM Events e1, Events e3
WHERE e1.ActivityLabel LIKE 'print%' AND e3.ActivityLabel LIKE 'print%'
      AND e1.Lifecycle = 'start' AND e3.Lifecycle = 'start'
      AND DATE(e1.Timestamp) = DATE(e3.Timestamp)  
      AND e1.Timestamp < e3.Timestamp
      AND ( 
        NOT EXISTS (
            SELECT * FROM Events e2
            WHERE e1.ProcessId = e2.ProcessId AND e1.TraceId = e2.TraceId
                  AND e2.ActivityLabel LIKE 'print%' 
                  AND e2.Lifecycle = 'complete'
                  AND e1.Timestamp < e2.Timestamp 
                  AND e2.Timestamp < e3.Timestamp
        ) )
      AND e1.Format <> e3.Format AND e1.Resource = e3.Resource;
\end{lstlisting}

\end{document}